\documentclass[layout=onecolumn, manuscript=letter]{achemso}
\usepackage[version=3]{mhchem} 
\usepackage[T1]{fontenc}       

\author{Peng-Lai Gong}
\affiliation{Department of Physics, South University of Science and Technology of China, Shenzhen 518055, China}
\alsoaffiliation{Department of Electronics and Tianjin Key Laboratory of Photo-Electronic Thin Film Device and Technology,  Nankai University, Tianjin 300071, China}
\altaffiliation{Contributed equally to this work}

\author{Bei Deng}
\affiliation{Department of Physics, South University of Science and Technology of China, Shenzhen 518055, China}
\altaffiliation{Contributed equally to this work}

\author{Liang-Feng Huang}
\affiliation{Department of Materials Science and Engineering, Northwestern University, Evanston, Illinois 60208, USA}

\author{Liang Hu}
\affiliation{Department of Physics, South University of Science and Technology of China, Shenzhen 518055, China}

\author{Wei-Chao Wang}
\affiliation{Department of Electronics and Tianjin Key Laboratory of Photo-Electronic Thin Film Device and Technology,  Nankai University, Tianjin 300071, China}

\author{Da-Yong Liu}
\email{dyliu@theory.issp.ac.cn}
\affiliation{Key Laboratory of Materials Physics, Institute of Solid State Physics,
             Chinese Academy of Sciences, P. O. Box 1129, Hefei 230031, China}

\author{Xing-Qiang Shi}
\email{shixq@sustc.edu.cn}
\affiliation{Department of Physics, South University of Science and Technology of China, Shenzhen 518055, China}

\author{Zhi Zeng}
\affiliation{Key Laboratory of Materials Physics, Institute of Solid State Physics,
 Chinese Academy of Sciences, P. O. Box 1129, Hefei 230031, China}
\alsoaffiliation{University of Science and Technology of China, Hefei 230026, China}

\author{Liang-Jian Zou}
\email{zou@theory.issp.ac.cn}
\affiliation{Key Laboratory of Materials Physics, Institute of Solid State Physics,
 Chinese Academy of Sciences, P. O. Box 1129, Hefei 230031, China}
\alsoaffiliation{University of Science and Technology of China, Hefei 230026, China}

\title{Robust and Pristine  Topological Dirac Semimetal Phase in Pressured Two-Dimensional Black Phosphorous}

\AbstractOn
\begin{document}
\begin{abstract}
  Very recently, in spite of various efforts in searching for two dimensional  topological Dirac semimetals (2D TDSMs)  in phosphorene, there remains a lack of experimentally efficient way to activate such phase transition and the underlying mechanism for the topological phase acquisition is still controversial. Here, from first-principles calculations in combination with a band-sorting technique based on $k\cdot p$ theory, a layer-pressure topological phase diagram is obtained and some of the controversies are clarified. We demonstrate that,
  compared with tuning by external electric-fields, strain or doping by adsorption, hydrostatic pressure can be an experimentally more feasible way to activate the topological phase transition for 2D TDSM acquisition in phosphorene. More importantly,
  the resultant TDSM state is a pristine phase  possessing a single pair of symmetry-protected Dirac cones {\it right} at the Fermi level, in startling contrast to the pressured {\it bulk} black phosphorous  where only a carrier-mixed Dirac state can be obtained.  We corroborate that the Dirac points are robust under external perturbation as long as the glide-plane symmetry  preserves.  Our findings provide a means to realize 2D pristine TDSM in a more achievable manner, which could be crucial in the realization of controllable TDSM states in phosphorene and related 2D materials.
\end{abstract}


\begin{tocentry}
 \includegraphics[width=0.90\textwidth]{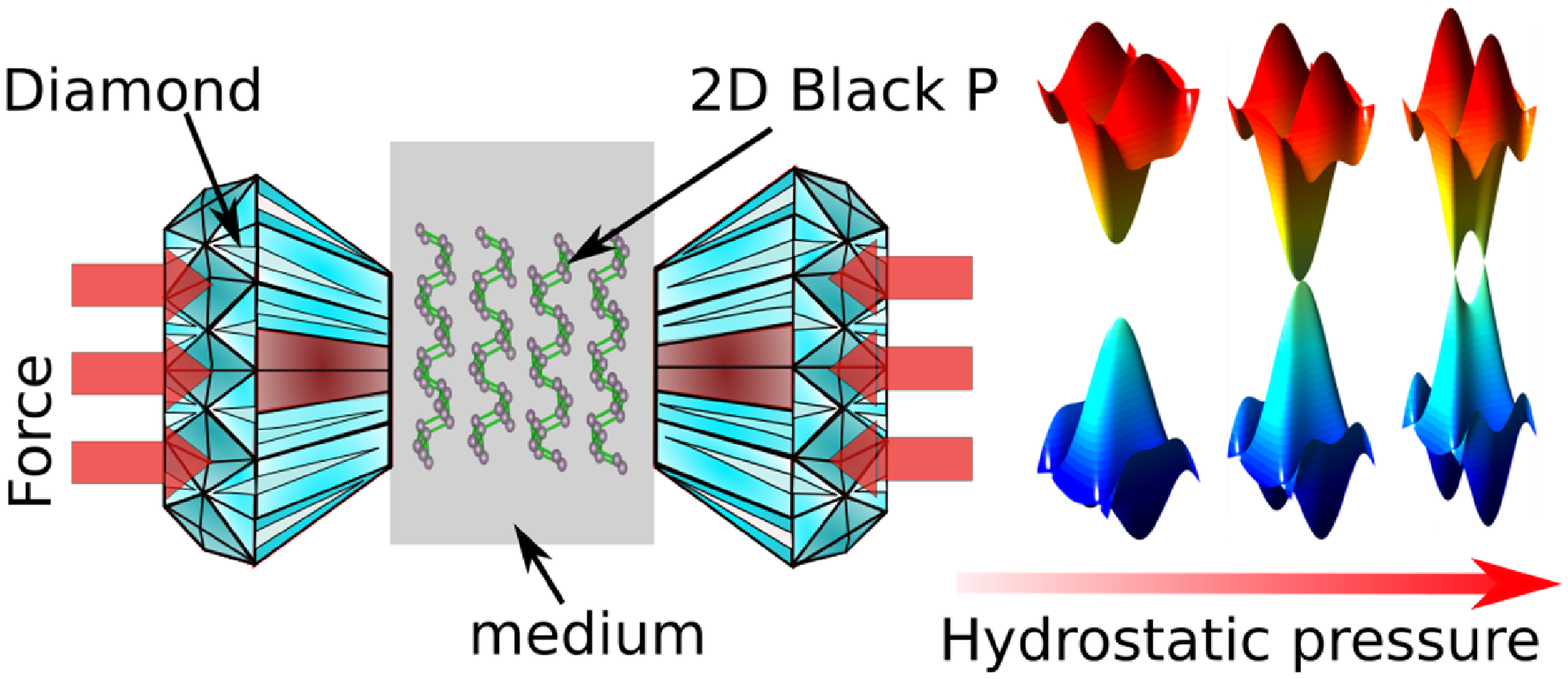}
\end{tocentry}

Two dimensional (2D) Dirac semimetals, in which Dirac points cross the Fermi level ($E_\text{F}$) being protected by nonsymmorphic crystal symmetries, was first proposed by Young and Kane in 2015 \cite{young2015dirac}. Very recent progress on the strain or electric-field modified phosphorene, as a representative of the rare candidates, has established a link between Dirac cones and the topological nature \cite{doh2016band, ghosh2016electric, fei2015topologically}. Given the experimental discovery of three dimensional ({\it 3D}) topological Dirac semimetals (Na$_3$Bi \cite{liu2014discovery}, Cd$_3$As$_2$ \cite{neupane2014observation}) and $\alpha$-Sn  on InSb(111) substrate \cite{xu2017elemental}, it is natural to ask whether the {\it 2D} topological Dirac semimetals (TDSMs) can also be realized in an experimentally  feasible manner.

A simple mechanism for a 2D TDSM phase acquisition has been proposed based on the Stark effect in phosphorene thin films by applying an external electric field \cite{ghosh2016electric, dolui2015quantum}. However, applying an exceptionally giant electric field on such a system is difficult to realize experimentally (the value of the field required is $\sim$0.5 V/\AA \/ for a four-layer phosphorene \cite{dolui2015quantum}). On the basis of the same mechanism, the experimentally observed Dirac semimetal state, from potassium doping of few-layer phosphorene, is actually an electron-doped TDSM  \cite{Kim20152D}. However, a pristine TDSM in 2D is highly desirable as it can be tuned to be topological insulators (TIs) or  Weyl semimetals by explicit breaking of symmetries. In this regard, in-plane strain has been proposed as a possible means  to induce Dirac cones in monolayer or bilayer phosphorene  \cite{lu2016multiple, Wang2015Strain, ding2014anisotropic}, whereas the critical strain (as large as $\sim$10\%  uniaxial  strain or  5\% biaxial strain based on the DFT-PBE level  \cite{Wang2015Strain}) is difficult to be experimentally realized, particularly for the biaxial strain in strong-anisotropic systems such as phosphorene.
Furthermore, such large uniaxial/biaxial strains could substantially lower the local symmetry of the system, resulting in large distortion and structural instability due to the accumulative strain energy.
Additionally, it remains conflicted  in literature about the true phase ({\it i.e.,} whether the resultant phase is a TI or a TDSM) in the strain-induced phase transition \cite{lu2016multiple, fei2015topologically, doh2016band}.
All the above issues need to be solved or clarified, and examined by  a more achievable way.

Apart from  uniaxial/biaxial strains, the hydrostatic pressures, which
to a larger extent preserve the crystal symmetry, and also could be free of the typical epitaxial-mismatch effect, have been
proved as a powerful tool to deal with both fundamental and practical issues.
Very recently, we have reported that {\it bulk} black phosphorous (BP) can convert from a normal insulator (NI) into a 3D Dirac semimetal under a certain hydrostatic pressure, both experimentally and theoretically, but only a carrier-mixed phase is acquired \cite{Xiang2015Pressure, Gong2016Hydrostatic}.  In experiments, hydrostatic pressures can also substantially modify the optical and vibrational properties of 2D systems like monolayer or few-layer MoS$_2$ \cite{Nayak2015Pressure, dou2016probing, dou2014tuning, li2015pressure, nayak2014pressure, yan2016interlayer}, WS$_2$ \cite{nayak2015pressureWS2}, and WSe$_2$ \cite{ye2016pressure}. Starting from this point, the few-layer phosphorene could be feasibly pressurized like other layered 2D systems.  Therefore, in this paper, from first-principles methods in conjunction with band-sorting technique based on  $k$$\cdot$$p$ theory (to solve the above mentioned TI or TDSM conflict), we study the layer-pressure topological phase diagram of few-layer phosphorene. We discovered that hydrostatic pressures can energetically drive the 2D phosphorene from NI to TDSM phase, with the topological-phase-transitions (TPTs)  depending on the number of layers.
The resultant 2D TDSM state is characterized to be a pristine phase as a consequence of the reduced dimensional effect in the vertical direction.
We demonstrate that the Dirac points are robust under external perturbation, including strain, pressure or electric field, as long as the glide-plane symmetry within each sublayer preserves.  

 In order to explore the evolutions of  electronic structures of few-layer phosphorene under increasing hydrostatic pressures, we first determine  their crystal structures  (under each pressure) by DFT calculations.
 We employed the Vienna Ab initio Simulation Package (VASP) \cite{hafner2008ab} with the projector augmented wave (PAW) method \cite{blochl1994projector,kresse1996efficient}.
 The interlayer vdW interaction is described by the optB88-vdW functional \cite{Klime2011optb88}.
 A previous study on few-layer MoS$_{2}$ has proposed an extraction method to obtain the 2D  structures under pressures \cite{fan2015electronic}.
Based on this method,  their results  reveal that the critical pressure for direct-to-indirect gap transition is $\sim$13 GPa, in qualitative agreement with the conclusion  drawn   from photoluminescence  experimental measurement of $\sim$16 GPa \cite{Nayak2015Pressure}.
The extraction method is a reasonable and practical strategy to mimic the real pressurized procedure for layered systems, because it makes the sample feel the pressure from the other parts of the bulk system, which could serve as a good approximation to the inert pressure-transmitting medium in experiments.
Meanwhile, the periodic potential field at the vdW boundary is not such strong that can change the geometric and electronic properties of the 2D BP system, as this is further verified by another method, namely, the medium method, in which the very realistic pressure-transmitting medium is considered.  We find based on a set of comparison test, that basically the two method will yield a same structure for the 2D BP under a given pressure (see Table S2), suggesting that the extraction method is reliable to the cases in this study. As it is computationally much more expensive to consider the medium in each configuration (more than 800 atoms in the whole system), especially for the cases with the inclusion of spin-orbital coupling (SOC) effect, here, we chose the extraction method as an appropriate approach to conduct the study. Within the extraction method, in our study the structure of {\it bulk} BP was first optimized under a specified hydrostatic pressure, and then these structural parameters were used to construct the $n-$layer phosphorene, with a vacuum space of 20 \AA \/ along the $z$ direction. Based on the constructed 2D structures, the electronic properties is calculated by hybrid functional of Heyd, Scuseria, and Ernzerhof (HSE06) \cite{heyd2006erratum} calculations  with the inclusion of SOC effect.

 \begin{figure}[!t]
 \noindent \begin{centering}
 \includegraphics[width=8.5cm]{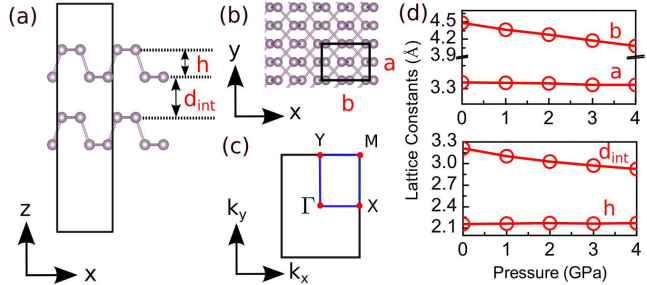}
 \par\end{centering}
 \caption{
  (a) Side- and (b) top-views of  the crystal structure of a 2-layer phosphorene; (c) 2D first Brillouin zone; (d) anisotropic variation of structural parameters of few-layer phosphorene under pressure.}
\label{Fig1-structures}
\end{figure}

  Few-layer BP is a layered material, in which each layer stacks together by van der Waals (vdW) interactions. Each layer consists of two sublayers and in turn forms a bulking honeycomb structure.  The structural model of  2-layer phosphorene, as a representative for a finite $n$-layer phosphorene ($n$ denotes the number of layers),  is shown in Figures \ref{Fig1-structures}a,b,   where the directions of lattice vectors $a$ and $b$ are along the zigzag (the $y$ axis) and the armchair (the $x$ axis) directions, respectively. The 2D rectangular Brillouin zone  of $n$-layer phosphorene is typified in Figure \ref{Fig1-structures}c.

  With increasing pressure, as shown in Figure \ref{Fig1-structures}d, the lattice constant $b$ (armchair direction) and  the interlayer spacing between two adjacent phosphorene layers ($d_\text{int}$) are significantly shortened (by 5.6\% and 8.8\% for $b$ and $d_\text{int}$  at 4.0 GPa), while $a$ (zigzag direction) and the sublayer distance within one single layer ($h$) remain almost unchanged.  The  thermodynamic and lattice-dynamic stabilities of few-layer phosphorene depend on their atomic and electronic properties, as well as the environment, such as pressure and temperature, {\it etc}. In the current work, the stabilities of few-layer phosphorene are in line with that of {\it bulk} BP. The  thermodynamic stability of the bulk BP is  considered through the enthalpy-pressure relationship.  We find that the critical pressure of thermodynamic stability $P_\text{T}$ for {\it bulk} BP with orthorhombic phase is 4.6 GPa (see Figure S2), very close to the experimental result (4.7 GPa) \cite{Akahama1997Raman}, which indicates a structural phase transition  from a A17 (orthorhombic) phase to a A7 (rhombohedral) phase  \cite{boulfelfel2012squeezing}. For few-layer phosphorene, a similar structural phase transition is found in accordance to our calculations because of the same mechanism of the structural phase transition as the {\it bulk} BP \cite{boulfelfel2012squeezing}.
  Phonon spectra are then calculated to judge whether or not the thermodynamically stable  structures (with $P\leq P_\text{T}$) are lattice-dynamic   stable. Our results show that all the pressured structures have no imaginary frequencies below $P_\text{T}$  \cite{Gong2016Hydrostatic}. The above results ensure that the pressured structures will  not experience a structural phase transition below $P_\text{T}$; and this is further confirmed by the medium method, showing that the 2D structure is indeed stable and does not favor reconstruction below $P_\text{T}$ (see Table S2 and Figure S1).

\begin{figure}[!t]
 \noindent \begin{centering}
 \includegraphics[width=8.5cm]{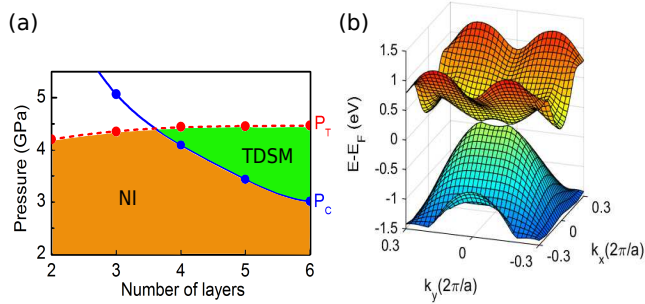}
 \par\end{centering}
  \caption{ Layer-pressure  topological phase diagram of  $n$-layer phosphorene. $P_\text{C}$ and  $P_\text{T}$  are the critical pressures for electronic phase transition and upper limit of  thermodynamic stability, respectively (see main text). NI denotes normal insulator (Figure 3a); and TDSM denotes a pure topological Dirac semimetal with Dirac points locating exactly at $E_\text{F}$ (Figures 3c,d). (b) Three dimensional band structure for the TDSM bands around $\Gamma$ point, showing a single pair of Dirac cones.}
 \label{phase}
\end{figure}

\begin{figure}[!t]
  \noindent \begin{centering}
  \includegraphics[width=8cm]{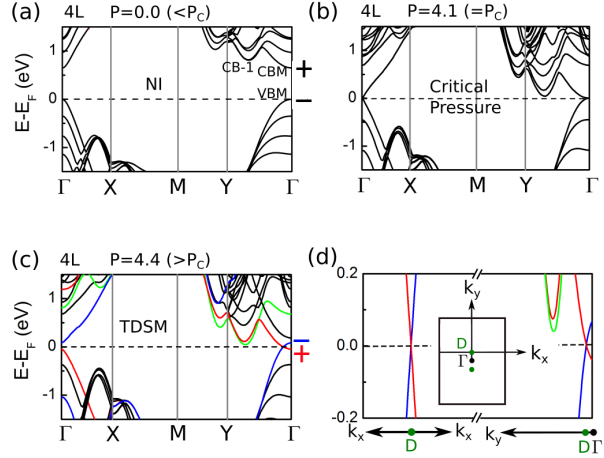}
   \par\end{centering}
   \caption{ Band structures of 4-layer phosphorene as a representative for the pressured structures  at (a) 0.0 GPa, (b) 4.1 GPa, (c) 4.4 GPa and (d) also 4.4 GPa, but around the Dirac point along the $k_x$ and $k_y$ directions, respectively. The SOC was also included in the calculation.
   The parity (even: +, odd: -) of the VBM and CBM at $\Gamma$ point is labeled in (a) and (c).
   Dirac point (denoted by green point, labeled with "D") was set as the starting point  in the band structure plotting in (d). A band-sorting method is used
   to sort the bands according to their symmetry (see text).
   }
  \label{Fig3-bands}
\end{figure}

{\it{Topological phase diagram.}}
 The layer-pressure  topological phase diagram  for few-layer phosphorene (below $P_\text{T}$, with consideration of SOC) is displayed in Figure \ref{phase}, where $P_\text{C}$ and  $P_\text{T}$  are the critical pressures for electronic and structural phase transitions, respectively. At $P_\text{C}$,  an electronic phase transition, from NI to TDSM, takes place (see Figure \ref{Fig3-bands}b). TDSM denotes a pristine topological Dirac semimetal phase with only Dirac fermions at $E_\text{F}$ (see Figure \ref{Fig3-bands}c). No TPTs are observed for $n=2$ and $3$ layers so they are still NIs below $P_\text{T}$ (dashed red line), this is because of their large band gaps ($\geq$ 0.75 eV) under zero pressure. With increasing number of $n$, their zero-pressure band gaps  gradually decrease, and a NI to TDSM transition happens to occur in the pressure range of $P_\text{C}$$<$$P$$<$$P_\text{T}$ for $n\geq4$. We should point out that, although the weak quantum confinement effect herein makes the band inversion occur at a higher pressure $P_\text{C}$ when referring to the case in the {\it bulk}, this effect could become less important in thicker layers (see Figure S3).

 Under pressure-free conditions, 4-layer phosphorene is a NI with a band gap of 0.63 eV as shown in Figure \ref{Fig3-bands}a. With increasing pressure, at $P=4.1$ GPa, the valance band maximum (VBM) and conduction band minimum (CBM) touch together at $\Gamma$ point of the Brillouin zone, as typified in Figure \ref{Fig3-bands}b.  When $P=4.4$ GPa, VBM and CBM with opposite parities have been inverted and a single pair of Dirac cones along $\Gamma-\text{Y}$ (and the opposite direction of $\Gamma-\text{Y}$) are formed (see Figure \ref{phase}b and Figures \ref{Fig3-bands}c, d), implying of a quasi-2D TDSM.

\begin{figure}[!b]
 \noindent \begin{centering}
 \includegraphics[width=8cm]{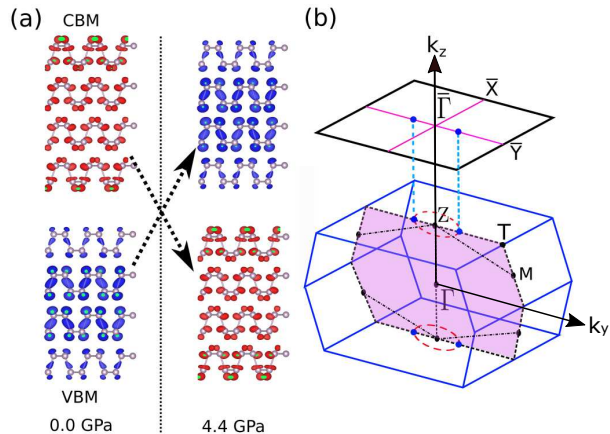}
 \par\end{centering}
 \caption{ (a) Real space charge distribution of the VBM and CBM at $\Gamma$ point under 0.0 and 4.4
  GPa, respectively. (b) Sketched distributions of the Dirac cones in the first
  Brillouin zone (BZ) of \emph{bulk} black phosphorus (bottom) and few-layer phosphorene (top) in their semimetal states with the consideration of SOC. The
   projected surface BZ is shown [in top of  (b)]. Red dashed line (in $\Gamma\text{ZT}$ plane) stands for the broken node-loop near the $E_\text{F}$, while blue solid points are the symmetry-protected Dirac points.}
 \label{Fig4-vbm}
\end{figure}

 By examining the real space distribution  of VBM and CBM at $\Gamma$ point,
 we find that the two bands are really inversed when $P$$>$$P_\text{C}$ (see Figure \ref{Fig4-vbm}a),  supporting the topological phase transition.
    The $Z_2$ topological invariant is calculated to further check whether the inverted band structure is topologically nontrivial or not.
  The calculation is carried out following the method developed by Fu and Kane \cite{fu2007topological}, based on the fact that inversion symmetry holds for all $n$-layer phosphorene studied here (Table S3). The $Z_2$ topological invariant is then  obtained from the parity of each pair of Kramers degeneracy occupied band at the time-reversal-invariant momenta (TRIM) points. As shown in Figure \ref{Fig1-structures}c, the BZ of $n$-layer phosphorene is a rectangle with four TRIM points: the $\Gamma$ point, the X point, the M point and the Y point. The $Z_2$ topological invariant is thus expressed by,
 \begin{equation}   
 \begin{aligned}
    \delta({\text K_i})&=\overset{N}{\underset{m=1}\Pi}\xi_{2m}^i ,   \\
       (-1)^\nu &=\overset{4}{\underset{i=1}\Pi}\delta({\text K_i})=\delta(\Gamma)\delta({\text X})\delta({\text M})\delta({\text Y}).
 \end{aligned}
 \end{equation}
 where $\delta$(K$_i$) stands for the product of parity eigenvalues at the TRIM points, $\xi=\pm 1$ are the parity eigenvalues and $N$ denotes the number of the degenerated occupied bands. Our results show that the inverted band structure has a non-zero integer $Z_2$ topological invariant ($\nu$=1), which ensures a nontrivial topological state.

 {\it{From 3D mixed Fermions to 2D pure Dirac Fermions.}}
  We have reported in an early recent work that {\it bulk} BP can convert from a NI into a 3D Dirac semimetal under the hydrostatic pressure, but only a carrier-mixed Dirac state was acquired \cite{Xiang2015Pressure,Gong2016Hydrostatic}.  The obtained 3D Dirac semimetal state displays a node-loop (red dashed circle) with continuous Dirac points around the Z point in BZ when SOC effect is not  considered (see Figure \ref{Fig4-vbm}b), which is in agreement with a previous theoretical study \cite{zhao2015topological}. In fact, only two pairs of them (on the loop  along $\text{Z}$-$\text{M}$ path) are exactly located  at $E_\text{F}$, while others are located within $\pm$0.15 eV around $E_\text{F}$ \cite{Gong2016Hydrostatic}. Even with consideration of SOC,  a single pair of Dirac cones along $\text{Z}$-$\text{T}$ path  cannot be opened up  due to the protection of glide-plane symmetry, while others on the loop are opened up with a gap ($\leq$10 meV). In the 2D case, the single pair of  Dirac cones are projected to the 2D BZ along the $\bar{{\Gamma}}$-$\bar{\text{Y}}$ path  (see Figure \ref{Fig4-vbm}b). For the pressured {\it bulk} BP, it only displays a mixture phase with the combined character of trival semimetals and topological Dirac semimetals (mTTDSM), which shows (hole-type and electron-type) Dirac fermions mixed with normal fermions \cite{Xiang2015Pressure,Gong2016Hydrostatic}. The mTTDSM phase in {\it bulk} BP  originates from the stronger anisotropic momentum and the charge compensation in $k_y$-$k_z$ plane. In contrast, the $n$-layer phosphorene is able to exhibit  a pure TDSM phase with a single pair of Dirac points locating exactly at $E_\text{F}$  under a certain pressure range, owing to the vanishment of anisotropic momentum $k_z$ from the reduced dimensional effect.



\begin{figure}[!b]
\noindent \begin{centering}
\includegraphics[width=8cm]{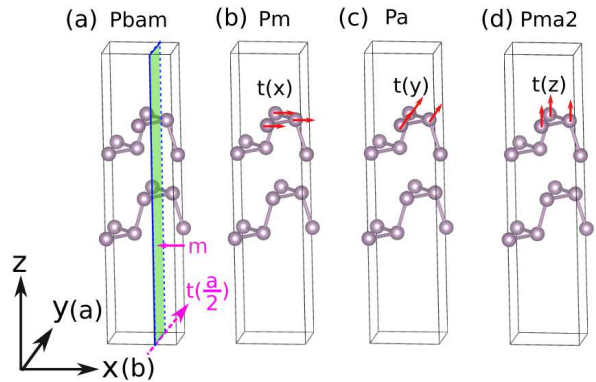}
\par\end{centering}
\caption{ Illustration of few-layer phosphorene (a) without deformations and (b-d) with particular deformations. Move the outmost P atoms  by 0.1 \AA  \/ along (b) $x$, (c) $y$ and (d) $z$ directions, respectively.  $t(x)$, $t(y)$ and $t(z)$ are translation vectors, and the space group for the structures  are also shown. The glide plane (green) in (a) is displayed, where operate $m$  and $t$ denote the mirror
and translation vector. }
\label{FigS4-txty}
\end{figure}

 {\it{Robust Dirac cones under external perturbation.}}
 In order to explicitly understand which  space group  elements  (glide plane and/or screw axis) protect the Dirac cones in few-layer phosphorene, we calculate the band structures of the few-layer structure with particular deformations above $P_\text{C}$. Three deformations are considered, by moving the outmost P atoms  by 0.1 \AA \/ along $x$, $y$ and $z$ directions, respectively (see Figure \ref{FigS4-txty}). Our results  show that translations along $y$ and  $z$ directions cannot result in Dirac cones opening (see Figure S4), because the YZ glide plane at the middle of the zigzag chain retains (see Table S4). Therefore, these Dirac cones are protected by the nonsymmorphic space symmetry (out-of-plane glide plane) in quasi-2D phosphorene.

Recently, numerous studies show contradiction about whether the strain-induced Dirac cones in monolayer or few-layer phosphorene can be opened up by SOC \cite{lu2016multiple, fei2015topologically, doh2016band}. To our knowledge, the DFT calculations usually sort different bands according to their magnitude discrepancy, which probably leads to artificial mini-gap in band structures (see Figure S5a). To understand the band structure of crystals, it is very important to sort the bands according to their eigenvector continuity. A band-sorting method, which is based on $k\cdot p$ theory \cite{huang2014correlation}, is designed here to sort the DFT wave functions (see the derivation steps in the Supporting Information), {\it i.e.},
   \begin{equation}   
 \begin{aligned}
 |\sum\limits_{i}{{{[c_{i}^{n_1, J_1}(k)]}^{*}}\cdot }c_{i}^{n_2, J_2}(k+{{\Delta }_{k}})|={{\delta }_{n_1, \hspace{0.1cm} n_2 }}\cdot {{\delta }_{J_1, \hspace{0.1cm} J_2 }}+O({{\Delta }_{k}}).
  \end{aligned}
 \end{equation}
 where  $n$, $J$ and $\Delta_{k}$ are the band index, total angular momentum and a small wave vector, respectively;  $c_{i}(k)$ is the coefficient of the $i$th atomic orbital in the wavefunction. The $k{\cdot}p$ theory has been applied to sort phonon bands, resulting in the discovery of various lattice dynamical mechanisms \cite{huang2014correlation,huang2015phonon}. Such basic algorithm is reformulated here to sort electronic bands, which can help us to efficiently distinguish mini-gap from band crossing. It is especially useful to investigate the complex electronic properties of topological materials. The resorted bands corroborate that the Dirac cones can not be opened up by SOC (see Figure S5b), further confirming that Dirac  points are protected by the intrinsic  symmetry ($i.e.$ glide plane). Our results are in good agreement with those derived from  the model methods \cite{doh2016band,fei2015topologically} with the consideration of the special symmetry elements.
 Recently, we notice that it remains in conflict between the theoretical models and DFT calculations about whether the Dirac cones induced by external electric field in few-layer phosphorene can be opened up by SOC \cite{Liu2015Switching,dolui2015quantum,doh2016band,ghosh2016electric}.
We then employ  our band-sorting methodology to deal with the electric field case. Our results explicitly show that the Dirac cones herein can not be opened up for a finite gap even with the largest field strength reported in literature \cite{Liu2015Switching,dolui2015quantum}, which can be understood by the fact that the glide-reflection symmetry retains  regardless of  the perpendicular electric fields \cite{doh2016band}.
Thus, here we elucidate that the Dirac cones are robust in respect to strain, pressure or external electric field, and this rationale is supported by a list of recent theoretial results and experimental observations \cite{doh2016band, ghosh2016electric, fei2015topologically, Gong2016Hydrostatic, Kim20152D, Xiang2015Pressure}.


In summary, we present in this work that a 2D pure topological Dirac semimetal phase (TDSM) can be feasibly and effectively achieved in few-layer phosphorene by adopting a moderate hydrostatic pressure. The electronic band structure of the 2D TDSM shows a single pair of Dirac points locating exactly at the Fermi level, which are protected by the nonsymmorphic space symmetry (glide plane)  and thus can not be opened up  by SOC.
 We also corroborate and clarify that these
 Dirac cones are robust under the external perturbation (including strain or external electric field), if not breaking  the glide-plane symmetry within each layer.
 The pressure-tunable topological properties of few-layer van der Waals materials may offer great flexibility in design and optimization of electronic and optoelectronic devices.

\begin{suppinfo}

This file contains five parts: computational details,  thermodynamic stability, quantum confinement effect, reduced dimensional effect and symmetry protected Dirac cones, and
sorting the band dispersions based on $k$$\cdot$ $p$ theory.
\end{suppinfo}

\begin{acknowledgement}
 The authors thank Dr. Rui Wang, Jin-Zhu Zhao, Xian-Long Wang and Jie Zhang for many helpful discussions on the subject. This work was supported by National key research and development program (Grant No. 2016YFB0901600), the NSF of China under Grant Nos. 11474145, 11334003, 11534010, the Nanshan Key Lab on Nonvolatile Memory Grant (KC2015ZDYF0003A), and the Special Program for Applied Research on Super Computation of the NSFC-Guangdong Joint Fund (the second phase) under Grant No. U1501501.
\end{acknowledgement}


\providecommand{\latin}[1]{#1}
\makeatletter
\providecommand{\doi}
  {\begingroup\let\do\@makeother\dospecials
  \catcode`\{=1 \catcode`\}=2\doi@aux}
\providecommand{\doi@aux}[1]{\endgroup\texttt{#1}}
\makeatother
\providecommand*\mcitethebibliography{\thebibliography}
\csname @ifundefined\endcsname{endmcitethebibliography}
  {\let\endmcitethebibliography\endthebibliography}{}

\newpage
\newpage

\end{document}


\section{S1. Computational details}
  First-principles calculations were performed using the projector-augmented-wave method \cite{blochl1994projector,kresse1996efficient} and a plane-wave basis set as implemented in the Vienna Ab initio Simulation Package (VASP) \cite{hafner2008ab}. In optimizing the geometry of the bulk BP under various pressures, van der Waals interactions were considered in terms of the optB88-vdW
  functional \cite{Klime2011optb88}. The energy cutoff for the plane-wave basis was set to 500 eV for all the calculations. The k-mesh of 13 $\times$ 11 $\times$ 4 and 13 $\times$ 11 in relaxation and electronic structure calculations were adopted to sample the 3D and 2D first Brillouin zone, respectively. The 2D geometries under specific pressures are obtained by extracting the structures  from the bulk BP under the same pressures. Based on these 2D structures, we carry out the band structures calculations with the hybrid functional of Heyd, Scuseria, and Ernzerhof (HSE06) \cite{heyd2006erratum} with the inclusion of spin-orbital coupling (SOC) effect. The  band-gap values of the 2D structures under zero-pressure, as well as additional theoretical result with structures by fully-relaxation method \cite{Qiao2014high} and the experimental result \cite{Cartz1979Effect} as for comparison, are listed in Table \ref{T-gap}. We find that the band-gap values drawn from the extraction method under zero-pressure are in good agreement with theoretical \cite{Qiao2014high} and experimental  \cite{Cartz1979Effect} results, with only tiny differences,
  indicating that our method is reliable in dealing with the anisotropic 2D structures.

  \begin{table*}[!htbp]
  \caption{Band-gap values from HSE06 for $n$-layer phosphorene ($n$=2-5) and \emph{bulk} BP under zero pressure. Results in previous theoretical and experimental studies are listed for comparison. }
  \begin{threeparttable}
  \begin{tabular}{cccccccc}
  \hline\hline
  &NL  \hspace{3cm}    & $E_g$ (eV)  \\
  \hline
  &2 \hspace{3cm}       & 0.95 (1.02\tnote{a} )     \\
  &3  \hspace{3cm}    & 0.75 (0.79\tnote{a} )     \\
  &4   \hspace{3cm}     & 0.63 (0.67\tnote{a} )     \\
  &5  \hspace{3cm}     & 0.58 (0.59\tnote{a} )     \\
  &bulk \hspace{3cm}  & 0.35 (0.36\tnote{a}, 0.31\tnote{b} )     \\
  \hline\hline
  \end{tabular}
   \begin{tablenotes}
        \footnotesize
        \item[a] Ref. [6], theory.    \\
        \item[b] Ref. [7], experiment.   \\
     \end{tablenotes}
  \end{threeparttable}
  \label{T-gap}
\end{table*}


{\bf Comparison of the standard method, the constraint method, the medium method and the extraction method  to pressurize 2D structures.}
In the framework of DFT, the hydrostatic pressure is obtained through stress tensor calculations. The macroscopic average tensor ($\sigma_{\alpha\beta}$) is the derivative of the total energy ($E_{total}$) with respect to the strain tensor ($\mu_{\alpha\beta}$) per unit volume,
 \begin{equation}   
 \begin{aligned}
    \sigma_{\alpha\beta}=-\frac{1}{\Omega}\frac{E_{total}}{\mu_{\alpha\beta}}
  \end{aligned}
 \end{equation}
 The matrices for 3D stress tensor $\sigma_{\alpha\beta}$ can be expressed as

 \begin{equation}
 \left(
 \begin{array}{ccc}
     \sigma_{11} &  \sigma_{12} &  \sigma_{13} \\
     \sigma_{21} &  \sigma_{22} &  \sigma_{23}  \\
    \sigma_{31} &  \sigma_{32} &  \sigma_{33}
 \end{array}
 \right)_.
 \end{equation}

 If the stress tensor is simply composed of a pressure that is the same in all directions,
 {\it i.e.,} three normal stresses meet $\sigma_{11}=\sigma_{22}=\sigma_{33}=\sigma_{H}$, then the shear stresses are zero.
  The hydrostatic stress ({\it i.e.} hydrostatic pressure) is expressed as
   \begin{equation}
 \left(
 \begin{array}{ccc}
     \sigma_{H}  &  0           &  0           \\
     0           &  \sigma_{H}  &  0           \\
     0           &  0           &  \sigma_{H}
 \end{array}
 \right)_.
 \end{equation}


 \begin{table}[!htbp]
  \caption{Structural information for $2$-layer phosphorene under various pressures on the basis of the extraction method and the medium method (in bracket).}
    \begin{tabular}{ccccccccc}
  \hline\hline
  P (GPa)   & $a$ (A)  & $b$ (eV) & $d_{int}$ (eV) & $R_1$ (eV) & $R_2$ (eV) & $\theta_1/\theta_1'$  & $\theta_2/\theta_2'$ (eV) \\
  \hline
  0   &3.34   &4.47  &3.19    &2.28   &2.24   &96.15    &102.42      \\
      &(3.33) &(4.47) &(3.17) &(2.28) &(2.24) &(95.91)  &(102.44) \\
 1    &3.34  &4.39  &3.09 &2.27 &2.24         &96.21    &101.85      \\
      &(3.32) &(4.39) &(3.07) &(2.27) &(2.24) &(95.81)  &(101.91) \\
 2    &3.34  &4.32  &3.02     &2.27   &2.24   &96.28     &101.33      \\
      &(3.31) &(4.32) &(2.98) &(2.27) &(2.23) &(96.07)   &(101.33) \\
 3    &3.33  &4.27  &2.96 &2.26 &2.23          &96.39    &101.03      \\
      &(3.32) &(4.27) &(2.94) &(2.26) &(2.23)  &(96.15)  &(100.95) \\
 4    &3.32  &4.21  &2.92     &2.26   &2.23    &96.48   &100.60      \\
      &(3.32) &(4.20) &(2.88) &(2.26) &(2.22) &(96.33)  &(100.44)\\
  \hline\hline
  \end{tabular}
  \label{T-medium}
   \end{table}

The above {\it standard method} has been successfully applied to study the structural properties of 3D materials under hydrostatic pressure, including anisotropic crystals (such as bulk black phosphorous \cite{fei2015topologically,Gong2016Hydrostatic} and the family of III-VI semiconductors \cite{ferlat2002ab}), with the obtained  results in agreement with experiments. \cite{kikegawa1983x, schwarz1992structural}.
  However, the standard method is not fitted to
a 2D system with a vacuum layer, because of the vanishment of confinement in the vacuum direction due to pressures and the ill-defined volume in Eq. (1) for a 2D system. Another method is called
 the {\it constraint method} that can avoid the vanishment of confinement by fixing the vacuum in the vertical direction of the simulation models, and the pressure on the vertical  direction is applied through shortening the interlayer distance $d_{int}$ (see Figure 1a for $d_{int}$). Although this method can find the structures with the matrix form of hydrostatic pressure like Eq. (2), it is an ideal model that ignores the weak interaction between 2D system and pressure-transmitting medium (such as a soft neon, liquid argon or silicone oil).
 Also, in the {\it constrained method} the total volume of the system is meaningless, leading to errors of stress tensors not only in vacuum direction, but also in  in-plane directions.
 The extraction method is a reasonable and practical strategy to mimic the real pressurized procedure for layered systems, because it could make the 2D sample feel the pressure from the other parts of the bulk system that serve as a good approximation to the inert pressure-transmitting medium in experiments. The similar pressure effect originates from the fact that the weak (vdW) interaction between the interlayers in such system (which is non-chemical-bonding interaction) can be regarded as a good simulation to the interaction between the pressure-transmitting medium and the few-layer material (which is also a non-chemical-bonding one). However, the 2D sample has to be extracted from the bulk since
the significant influence of the dimensional and size effects on the electronic structure.

\begin{figure}[!t]
 \noindent \begin{centering}
 \includegraphics[width=16cm]{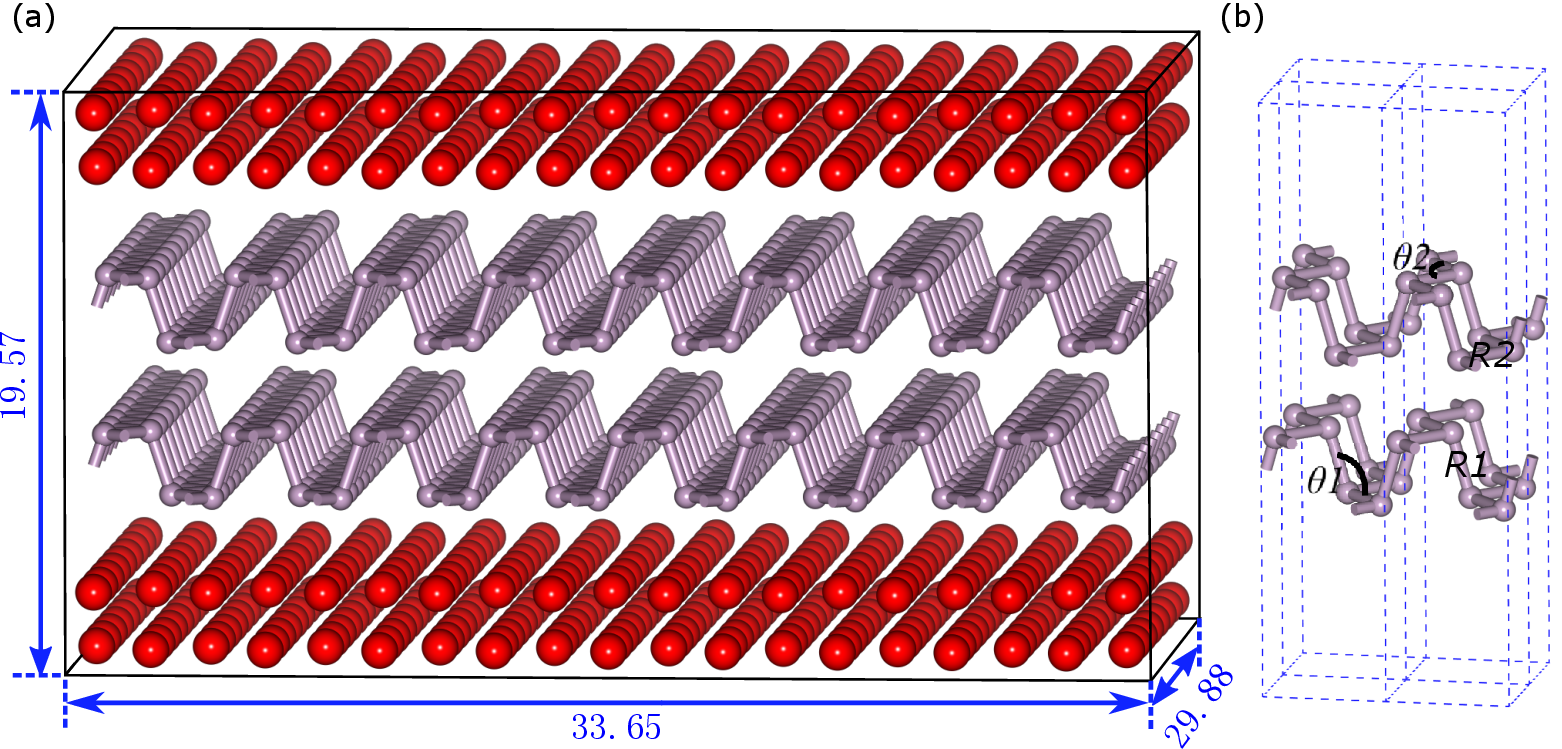}
 \par\end{centering}
 \caption{ (a) Optimized structure of 2-layer phosphorene at 4.0 GPa with the Ne atoms considered as the pressure-transmitting medium. For this calculation, a superlattice
combing an orthogonal supercell (8$\times$9) of Ne and a 9$\times$8 supercell of 2-layer phosphorene is constructed. The lattice constants $a$ and $b$ of the superlattice are kept fixed to maintain a uniform in-plane stress corresponding to such pressure, while the $c$ axis is allowed to relax to vary the out-of-plane stress, in order to achieve a hydrostatic pressure. Here, the lattice constants  $a$ and $b$ of the superlattice are derived from that of pure BP and Ne calculated under the same pressure. Note here that due to the lattice mismatch of BP and Ne under pressures, consider the case of 4.0 GPa for example, an extremely large superlattice with more than 800 atoms, which consists of 8$\times$9 supercell of Ne and 9$\times$8 supercell of phosphorene, is employed in order to simulate the uniform in-plane stress. During the calculation, all the atoms are allowed to relax until the calculated Hellmann-Feynman forces are less than 0.03 eV/\AA.
 (b) Structural parameters of phosphorene ($\theta1$, $\theta2$, $R1$ and $R2$).}
  \label{medium-str}
\end{figure}

Here, we also introduce another method, namely, the {\it medium method}, in which the realistic pressure-transmitting medium is considered. In this method, Ne atoms are chosen as a prototypical pressure medium to fill the vacuum layer of 2D BP. As a result, the overall system turns to be 3D, to which the standard method could be again applicable.
This method can be regarded as a means most approaching the experimental condition, and hence also a good benchmark to check the reliability of the extraction method. In this regard,
our results suggest that the 2D structure has merely negligible difference in these two methods (see Table S2 and Figure S1), revealing that the extraction method is  reliable in our study. Also, no structural reconstruction was observed below $P_\text{T}$, in either the two methods.
Although the medium method is considered as the best one to mimic the real experiments, the limited advantage as above can not compete with its huge computational cost (i.e., simulation with more than 800 atoms). Therefore, in our work, the extraction method is selected as the major approach for the thorough calculations.

In a word, the extraction method with lower calculation cost  is
an appropriate strategy to mimic the  pressurized layer systems.
At last, we should point out here that, the geometry of the extracted 2D structure, which is adopted for the study of the electronic and topological properties, is kept constrained after extraction, since the matrix of the stress tensor for an extracted structure is physically meaningless.

\begin{figure}[!b]
 \noindent \begin{centering}
 \includegraphics[width=10cm]{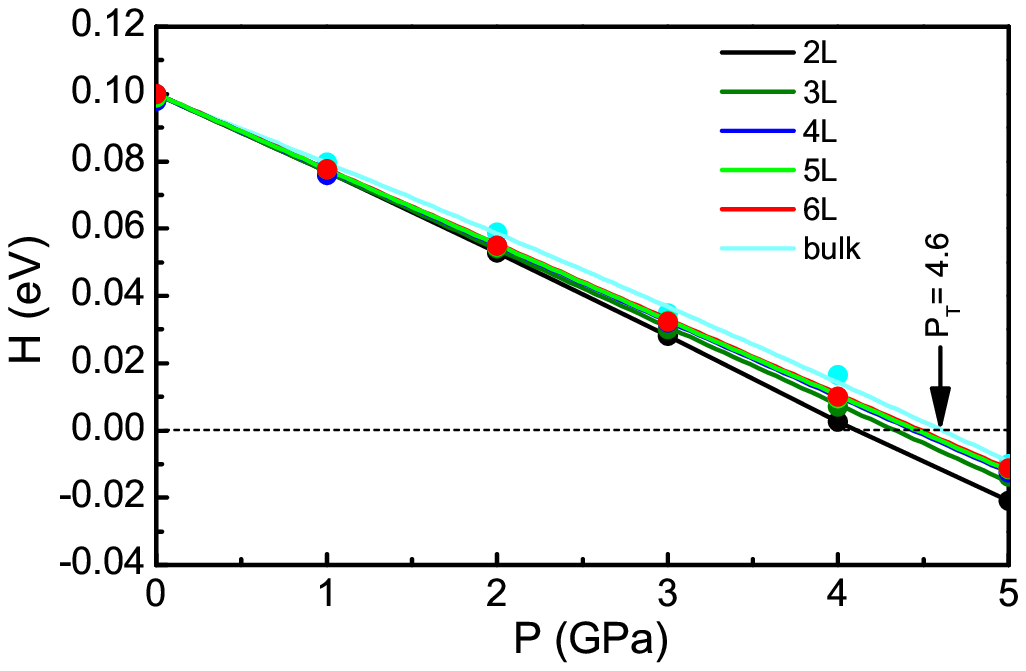}
 \par\end{centering}
 \caption{Relationship of enthalpy and pressure
     for A17 and A7 phases. A17 and A7 phases are orthorhombic and rhombohedral structures of bulk and $n$-layer BP in different regions of pressure. The enthalpy of the A17 phase is set to be zero as a reference to the A7 phase.}
 \label{HP}
\end{figure}

\section{S2. Thermodynamic stability}
As the most stable allotrope of phosphorus, bulk black
phosphorus exhibits three different phases under moderately
high hydrostatic pressures: the orthorhombic phase (A17) with
wrinkled hexagons, the graphene-like rhombohedral phase (A7)
with hexagonal lattice  for the pressure $P$ $>$ 4.7 GPa,
and the simple cubic phase for $P$ $>$11 GPa \cite{Akahama1997Raman}.
The thermodynamic stability for the bulk BP can be employed
to study the structure phase transition, through the enthalpy-pressure relationship for different phases. We calculate the enthalpy-pressure relationship for A17 and A7 phases of the bulk BP  in the region of $0 \leq P \leq 5$ GPa in the HSE06 level. As shown in Figure \ref{HP}, we find the critical pressure for structural phase transition  is 4.6 GPa, in good agreement with the experimental result \cite{Akahama1997Raman}. For 2D phosphorus, an equimolar thickness \cite{marks2016nanoparticle} (here the distance between the outmost P atoms plus the distance between two adjacent phosphorene layers) is introduced to calculate the volume and then the corresponding enthalpy ($H=E_{total}+PV$).

\section{S3. Quantum confinement effect}
\begin{figure}[!b]
\noindent \begin{centering}
\includegraphics[width=9cm]{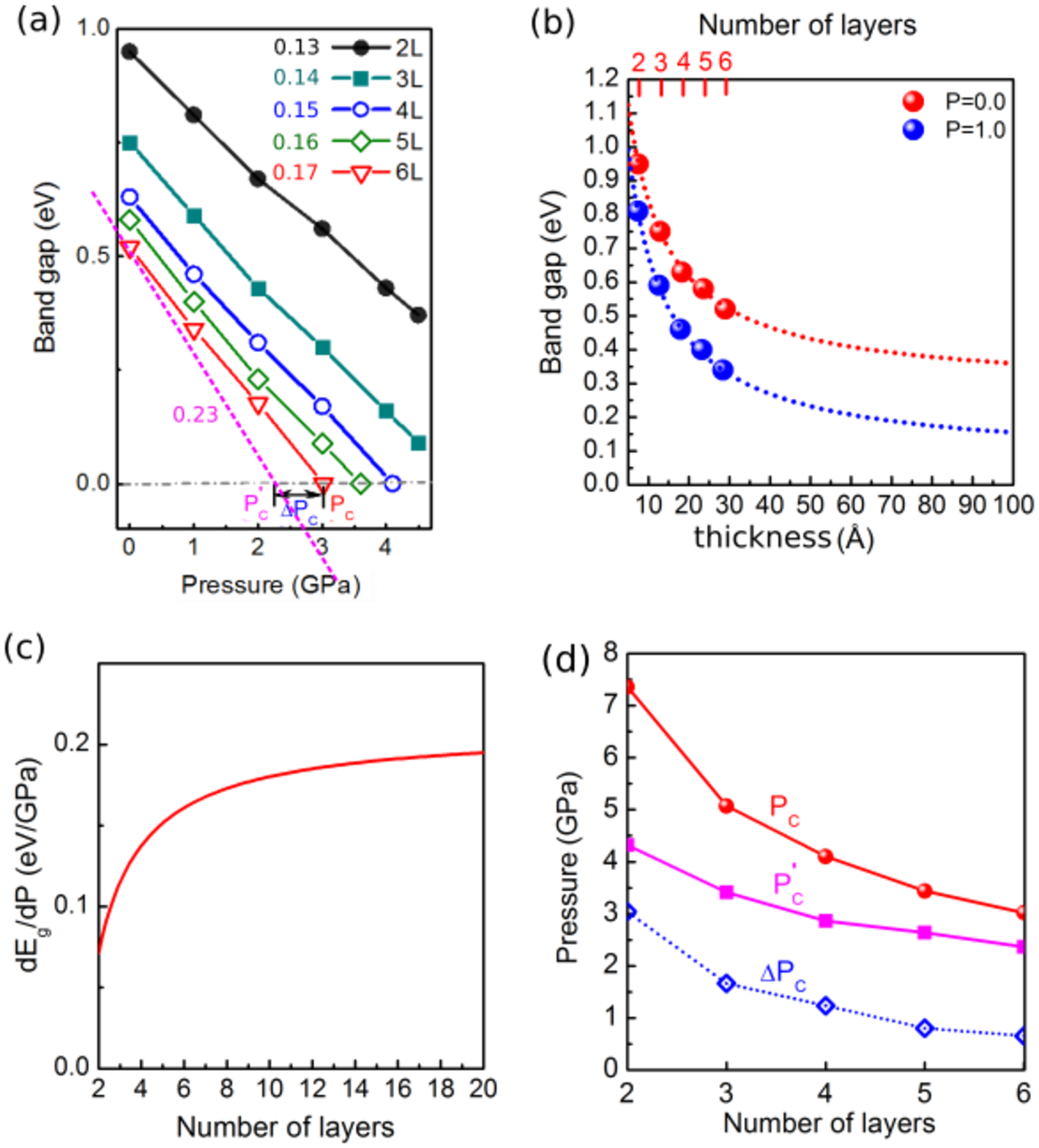}
\par\end{centering}
\caption{ (a) Band gaps of $n$-layer phosphorene as a function of pressure. The slope $dE_g/dP$ in units of eV/GPa is given near the legend; $P_C$ and $P_{C}'$ denotes the critical pressures with the inclusion and exclusion of QCE, and we define $\triangle P$ = $P_{C}$-$P_{C}'$.
  The pink dashed line for {\it bulk} BP without QCE is also shown.
(b) Band-gap values at $P$=0.0 (red) and $P$=1.0 GPa (blue) as
   a function of the confinement size. The thickness and its corresponding layers for $P$=0.0 GPa are illustrated. The fitted curves (red and blue dashed lines) are
   obtained according to the formula $E_g$=a/(t+b)+c, where a, b and c are fitted parameters. (c) The absolute value of the slope $k=dE_g$/$dP$  as a function of the number of layers.  (d) $P_C$, $P_{C}'$  and $\triangle P$ as a function of the number of layers.
   }
\label{QCE}
\end{figure}

As shown in Figure \ref{QCE}a, the band gap $E_\text{g}$  decreases linearly with the increasing pressure, and eventually closes at a critical pressure $P_{C}$ for $n\geq 4$ cases. Moreover, $P_\text{C}$ also decreases with the increase of thickness, indicating that the TDSM phases could be more realizable  in thicker layers. The absolute value of the slope $dE_\text{g}/dP$ is calculated to be 0.23 eV/GPa for {\it bulk} BP (the dashed line in Figure \ref{QCE}a), while it is 0.17, 0.16, and 0.15 eV/GPa for 6-, 5- and 4-layer phosphorene, respectively (the solid lines in Figure \ref{QCE}a).
 The decreased $dE_\text{g}/dP$ here leads to the band-gap closing at a slower rate in few-layer phosphorene than that happened in the bulk.
 The  decreased $dE_\text{g}/dP$ is supposed to be attributed  to the weak quantum confinement effect (QCE) \cite{yang2007quasiparticle,ghosh2016electric} in few-layer phosphorene.

 For the confined nanosystems, such as graphene nanoribbons (GNRs), the band-gap values shows clearly size  dependence because of the quantum confinement effect (QCE) \cite{yang2007quasiparticle}. Under a hard-wall boundary condition and considering the effective width of GNRs, the formula $E_g$=$a/(w+w_0+\delta_0)$ can well describe the relation of band gap ($E_g$) and the width of 1D GNRs ($w$). Here, for the 2D few-layer phosphorene, the thickness ($t$)  confines the system and influences the band gap.
 It should have a finite band gap when the thickness is approaching the bulk below the critical pressure of 1.2 GPa. Therefore, we  use a modified formula
\begin{equation}
  E_g=\frac{a}{t+b}+c,
\end{equation}
to fit the band gap values in Figure \ref{QCE}a, and then get  the fitted results
\begin{equation}
\begin{aligned}
  E_g^{P=0.0}=\frac{8.37}{t^{P=0.0}+4.91}+0.28, \\
  E_g^{P=1.0}=\frac{8.90}{t^{P=1.0}+4.66}+0.07,
\end{aligned}
\end{equation}
Our results show that the size dependence of band gaps for the pressured few-layer phosphorene (in case of $P$=0.0 and 1.0 GPa) can be well
expressed by an inverse relation Eq. (5).
Based on the structural information under  $P=0.0$ and $1.0$ GPa,
we obtain the relationships between $t$ and $n$:
\begin{equation}
\begin{aligned}
  t^{P=0.0}=5.37\cdot n-3.21, \\
  t^{P=1.0}=5.26\cdot n-3.10,
\end{aligned}
\end{equation}
where $n$ is the number of layers ($n\geq 2$). The slope $k$=$dE_g$/$dP$  can be expressed by
\begin{equation}
\begin{aligned}
 k  &= \mid\frac{E_g^{P=0.0}-E_g^{P=1.0}}{\triangle P} \mid \\
   & = \frac{8.37}{t^{P=0.0}+4.91}-\frac{8.90}{t^{P=1.0}+4.66}+0.21,
  \end{aligned}
\end{equation}
Combining Eqs. (6)  and (7),  $k$ can be directly relative to $n$:
\begin{equation}
\begin{aligned}
 k  = \frac{8.37}{5.37\cdot n+1.70}- \frac{8.90}{5.26\cdot n+1.56}+0.21.
  \end{aligned}
\end{equation}

In a given pressure range (0 to 1 GPa, for instance), the band gap change rate $dE_\text{g}/dP$ shows a clear size-dependence on the thickness along the $z$ direction, showing that in addition to the QCE dependence of the band gap $E_\text{g}$ (larger $E_\text{g}$ for thinner layers), the $dE_\text{g}/dP$ is also QCE-dependent, namely, smaller $dE_\text{g}/dP$ for thinner layers (Figure \ref{QCE}b). As shown in Figure \ref{QCE}c, the band gap change rate $dE_g$/$dP$ increases with increasing layers, owing to the decreased QCE in thicker layers.  The decreased slope $k=dE_g$/$dP$ in few-layer phosphorene resists the band-gap reduction, which, together with the QCE-induced larger gap,
leads to the band-gap closing at  a higher $P_C$ than expected.
Here, we define the difference between the critical pressures
with the inclusion and exclusion of QCE as $\triangle P$ =
$P_{C}$-$P_{C}'$, which can reflect the QCE on the ideal critical pressure $P_{C}'$ for topological phase transitions. From Figure \ref{QCE}d, we find that $\triangle P$ decreases with the increasing of layers, indicating the QCE can be significantly reduced in thicker layers. For example, the contributions of $\triangle P$ to $P_{C}'$ are 50\% for 4 layers and 33\% for 6 layers, respectively.

\section{S4. Reduced dimensional effect and symmetry protected Dirac cones}
\begin{figure*}[!t]
 \noindent \begin{centering}
 \includegraphics[width=12cm]{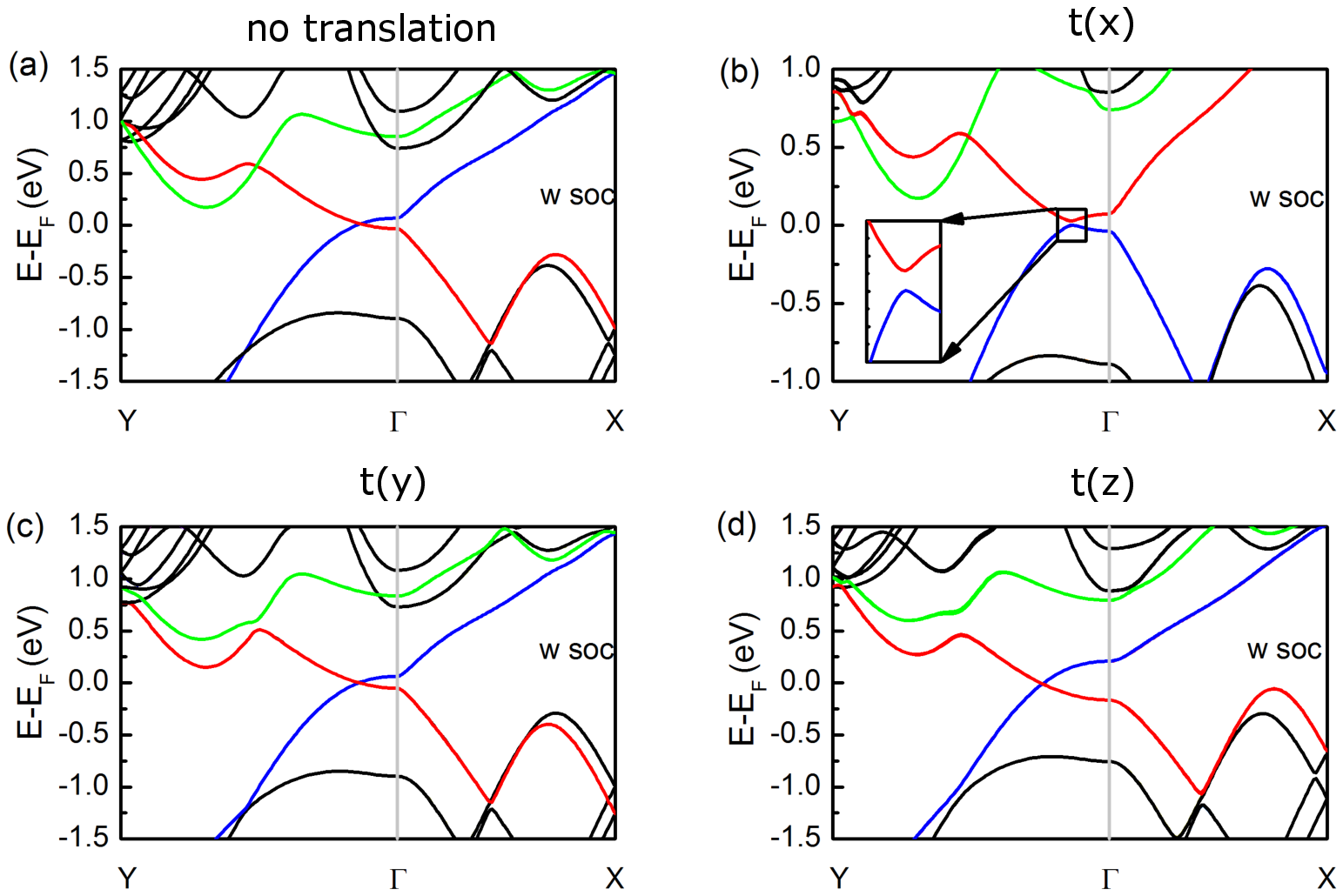}
 \par\end{centering}
\caption{ The evolution of Dirac cones in few-layer phosphorene with particular deformations (illustrated in Figure 5 in the main text).
 The band structures with and without SOC (not shown) are almost the same for each deformed structure.
 We can see explicitly  that the deformation with $t(y)$ and $t(z)$ cannot open the Dirac point, while that with $t(x)$ open a tiny gap (about 29 meV) at the Dirac point.}
\label{dirac-def}
 \end{figure*}

 Some symmetry elements for few-layer phosphorene and \emph{bulk} BP are listed in Table \ref{bulk-sym}. Odd-layer, even-layer phosphorene and bulk BP  preserve glide planes and screw axis, and their space groups do not change under hydrostatic pressures.  Some symmetry elements for 4-layer phosphorene, with particular deformations under 4.4 GPa (Figure 5), are listed in Table \ref{def-sym}. We find that a finite gap can be observed in the vicinity of  the Dirac points with $t(x)$ deformation due to breaking the YZ glide plane, but the Dirac states can not be opened up with $t(y)$ and $t(z)$ deformations (Figure \ref{dirac-def}) where Dirac points are protected by the nonsymmorphic space symmetry ({\it i.e.} YZ glide plane at the middle of the zigzag chain) of quasi-two-dimensional phosphorene.

\begin{table*}[!htbp]
\caption{Some symmetry elements for few-layer phosphorene and \emph{bulk} BP.}
\begin{tabular}{ccccccc}
  \hline\hline
     \hspace{1.5cm}      &glide plane \hspace{-0.2cm}  &screw axis \hspace{0.1cm} &inversion \hspace{0.2cm}   &space group \hspace{0.2cm}   \\
  \hline
  bulk       &YZ, XY, XZ  &X,Y,Z  \hspace{0.2cm} &YES   \hspace{0.2cm} &($D_{2h}^{18}$, Cmca)$^a$          \\
  \hline
  odd layer  &YZ, XY      &X  \hspace{0.2cm} &YES   \hspace{0.2cm}    &($D_{2h}^7$, Pmna)$^a$           \\
  \hline
  even layer &YZ, XY      &X \hspace{0.2cm}  &YES    \hspace{0.2cm}   &($D_{2h}^{11}$, Pbam)$^a$            \\
  \hline\hline
\end{tabular}
\\
 \textsuperscript{\normalsize $^a$Schoenflies symbol, Hermann-Mauguin symbol.}
 \label{bulk-sym}
\end{table*}

\begin{table*}[!htpb]
\caption{ Some symmetry elements for few-layer phosphorene  after particular deformations.}
\begin{tabular}{cccccccc}
 \hline\hline
    \hspace{0.0cm}      &glide plane \hspace{-0.2cm}  &screw axis \hspace{0.1cm} &inversion \hspace{0.2cm}   &space group \hspace{0.2cm}  \hspace{0.2cm} &Dirac points  \\
 \hline
  no translation       &YZ, XY  &X  \hspace{0.2cm} &YES   \hspace{0.2cm}    &($D_{2h}^{11}$, Pbam)$^a$          &YES \\
  \hline
  t(x)        &NO      &NO \hspace{0.2cm}  &NO    \hspace{0.2cm}   &($C_s^1$, Pm)$^a$             &NO \\
  \hline
  t(y)                 & YZ     &Z  \hspace{0.2cm} &NO   \hspace{0.2cm}    &($C_{s}^2$, Pa)$^a$            &YES\\
  \hline
  t(z)              &YZ     &Y \hspace{0.2cm}  &NO    \hspace{0.2cm}   &($C_{2v}^4$, Pma2)$^a$       &YES \\
  \hline\hline
\end{tabular}
\\
 \textsuperscript{\normalsize $^a$Schoenflies symbol, Hermann-Mauguin symbol.}
  \label{def-sym}
\end{table*}

\section{S5. Sorting the band dispersions based on $k\cdot p$ theory}
Quasiparticles in materials (e.g., electrons, phonons, and magnons) may have complex band dispersions. Sorting the bands according to their eigenvectors, instead of the way to their magnitudinal orders as widely applied  in most DFT codes, is much helpful to understand the band properties and reveal
 important underlying physics.

   The $k\cdot p$ for electronic wavefunctions can be expressed as
   \begin{equation}   
\begin{aligned}
   |<{{\Psi }_{n_1, J_1}}(k)|{{\Psi }_{n_2, J_2}}(k+{{\Delta }_{k}})>|=&{{\delta }_{n_1 ,n_2 }}\cdot {{\delta }_{J_1, \hspace{0.1cm} J_2 }}+O({{\Delta }_{k}})\in (0,1),
 \end{aligned}
  \end{equation}
 where  $n$, $J$ and $\Delta_{k}$ are the band index, total angular momentum (good quantum number in the inclusion of SOC) and a small wave vector, respectively.
  The electronic wavefunction can be expressed using a linear combination of atomic orbitals (LCAO)
  \begin{equation}   
 \begin{aligned}
    {{\Psi }_{n, J}}(k)=\sum\limits_{i}{c_{i}^{n, J}(k){{\phi }_{i}}},
 \end{aligned}
 \end{equation}
 where $c_{i}(k)$ is  the projection coefficient of the  the $i$th atomic orbital in the wavefunction $\Psi_{n, J}(k)$. When the used atomic orbitals construct an orthonormal basis set, there will be
 \begin{equation}   
 \begin{aligned}
     <{{\phi }_{i}}|{{\phi }_{j}}>={{\delta }_{i,j}},
 \end{aligned}
 \end{equation}
 Combining  Eqs. (9-11), we can obtain
  \begin{equation}   
 \begin{aligned}
 |\sum\limits_{i}{{{[c_{i}^{n_1, J_1}(k)]}^{*}}\cdot }c_{i}^{n_2, J_2}(k+{{\Delta }_{k}})|={{\delta }_{n_1, \hspace{0.1cm} n_2 }}\cdot {{\delta }_{J_1, \hspace{0.1cm} J_2 }}+O({{\Delta }_{k}}).
  \end{aligned}
 \end{equation}
 which can be used as an efficient and strict criterion to sort bands.

   We  used our method to check the bands nature of the  present Dirac semimetals, such as Na$_3$Bi \cite{wang2012dirac}, Cd$_3$As$_2$ \cite{wang2013three} and CaTe \cite{du2016cate}. Our results successfully confirm that the band crossing in these systems can not be opened up by SOC. In this regard,  we suggest that, usually  a mini-gap with the DFT value of smaller than 10 meV should be further checked in accordance to the related bands symmetry, as we have proposed in this method.

\begin{figure*}[!htpb]
 \noindent \begin{centering}
 \includegraphics[width=12cm]{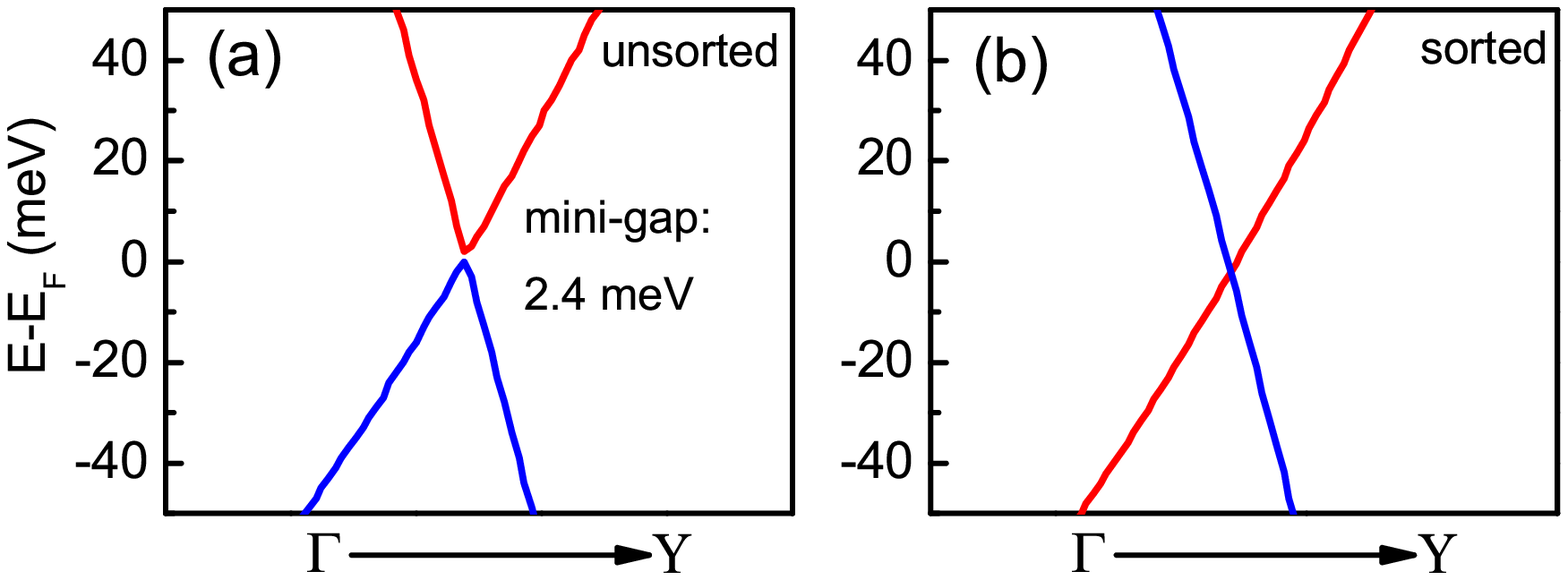}
 \par\end{centering}
\caption{ Band structures of 4-layer phosphorene at 4.4 GPa, (a) without and (b) with the band-sorting technique, respectively. It shows a mini-gap of 2.4 meV for the unsorted, and band crossing for the sorted one.}
\label{minigap}
 \end{figure*}

\newpage


\providecommand{\latin}[1]{#1}
\makeatletter
\providecommand{\doi}
  {\begingroup\let\do\@makeother\dospecials
  \catcode`\{=1 \catcode`\}=2\doi@aux}
\providecommand{\doi@aux}[1]{\endgroup\texttt{#1}}
\makeatother
\providecommand*\mcitethebibliography{\thebibliography}
\csname @ifundefined\endcsname{endmcitethebibliography}
  {\let\endmcitethebibliography\endthebibliography}{}


